\documentclass[prl,aps,twocolumn,showpacs,superscriptaddress,raggedbottom]{revtex4}
\usepackage{graphicx,bm}
\usepackage{dcolumn}
\usepackage{amsmath}

\usepackage{color}

\input{epsf}

\def\eq#1{Eq.~(\ref{#1})}
\def\fig#1{Fig.~\ref{#1}}

\begin{document}

\title{Strain-Induced Coupling of Spin Current to Nanomechanical Oscillations}

\author{A.~G. Mal'shukov}
\affiliation{Institute of Spectroscopy, Russian Academy of Science,
142190, Troitsk, Moscow oblast, Russia}
\author{C.~S. Tang}
\affiliation{Physics Division,
National Center for Theoretical Sciences, P.O. Box 2-131, Hsinchu
30013, Taiwan}
\author{C.~S. Chu}
\affiliation{Department of Electrophysics, National Chiao Tung
University, Hsinchu 30010, Taiwan}
\author{K.~A. Chao}
\affiliation{Solid State Theory Division, Department of Physics,
Lund University, S-22362 Lund, Sweden}

\begin{abstract}
We propose a setup which allows to couple the electron spin degree
of freedom to the mechanical motions of a nanomechanical system not
involving any of the ferromagnetic components. The proposed method
employs the strain-induced spin-orbit interaction of electrons in
narrow gap semiconductors. We have shown how this method can be used
for detection and manipulation of the spin flow through a suspended
rod in a nanomechanical device.
\end{abstract}


\maketitle

An ability to control the spin transport in semiconductors is a key
problem to be solved towards implementation of semiconductor
spintronics into quantum information
processing~\cite{Loss02,Wolf01,Rev.Mod.Phys}. Many methods have been
proposed to achieve control of the electron spin degree of freedom
using magnetic materials, external magnetic fields and optical
excitation (for a review see Ref.~\cite{Rev.Mod.Phys}). Other
promising ideas involve the intrinsic spin-orbit interaction (SOI)
in narrow gap semiconductors to manipulate the spin by means of
electric fields~\cite{Sinova03} and electric
gates~\cite{Datta,Malsh,Tang}. Recently, some of these ideas have
been experimentally confirmed~\cite{Wunderlich,Kato}.

In semiconductors the spin-orbit effect appears as an interaction of
the electron spin with an effective magnetic field whose direction
and magnitude depend on the electron momentum. A specific form of
this dependence is determined by the crystal symmetry, as well as by
the symmetry of the potential energy profile in heterostructures. In
strained semiconductors new components of the effective magnetic
field appear due to violation of the local crystal
symmetry~\cite{Soibystrain}. The effect of the strain-induced SOI on
spin transport was spectacularly demonstrated by Kato \emph{et. al.}
in their Faraday rotation experiment~\cite{Kato}. An interesting
property of the strain-induced SOI is that the strain can be
associated with mechanical motion of the solid, in particular, with
oscillations in nanomechanical systems (NMS), in such a way making
possible the spin-orbit coupling of the electron spin to
nanomechanical oscillations. At the same time a big progress in
fabricating various NMS~\cite{Roukes} allows one to reach the
required parameter range to observe subtle effects produced by such
a coupling.
\begin{figure}[b]
\begin{center}\leavevmode
\includegraphics[width=.5 \textwidth,angle=0]{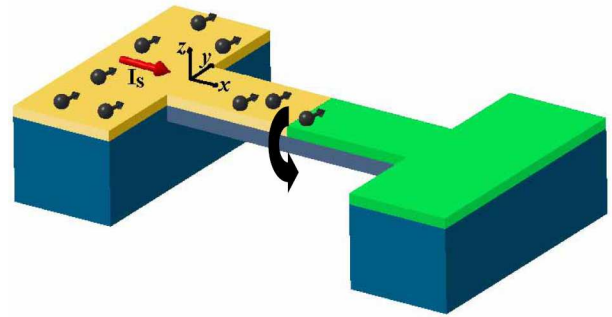}
\caption{Schematic illustration of electromechanical spin current
detector, containing a suspended semiconductor-metal (S-M)
rectangular rod atop an insulating substrate (blue). A spin current
is injected from the left semiconductor reservoir (yellow) and then
diffuses toward the metallic film (green). While passing through the
semiconductor film, the spin current induces torque shown by the
black arrow.}\label{system}
\end{center}
\end{figure}

In this Letter we will consider NMS in the form of a suspended beam
with a doped semiconductor film epitaxially grown on its surface
(see \fig{system}).  An analysis of the SOI in this system shows
that the flexural and torsion vibrational modes couple most
efficiently to the electron spin. As a simple example, we will focus
on the torsion mode. The strain associated with torsion produces the
spin-orbit field which is linear with respect to the electron
momentum and is directed perpendicular to it. This field varies in
time and space according to respective variations of the torsion
strain. Due to the linear dependence on the momentum, the SOI looks
precisely as interaction with the spin dependent electromagnetic
vector potential. An immediate result of this analogy is that the
time-dependent torsion gives rise to a motive force on electrons.
Such a force, however, acts in different directions on particles
with oppositely oriented spins, inducing thus the spin current in
the electron gas. The physics of this phenomenon is very similar to
the spin current generation under time-dependent Rashba SOI, where
the time dependence of the SOI coupling parameter is provided by the
gate voltage variations \cite{Malsh}. In the present work we will
focus, however, on the inverse effect. Due to the SOI coupling, the
spin current flowing through the beam is expected to create a
mechanical torsion. The torque effect on NMS due to spin flow has
been previously predicted by Mohanty \emph{et. al.}~\cite{Mohanty}
for a different physical realization, where the torque has been
created by spin flips at the nonmagnetic-ferromagnetic interface.
They also suggested an experimental setup to measure such a small
torque. As it will be shown below, the torque due to the strain
induced SOI can be large enough to be measured using the
experimental setup proposed in Ref.~\cite{Mohanty}. Besides this
method, other sensitive techniques for displacement measurements can
be employed \cite{measurements}.

The system under consideration is a rectangular beam of the total
length $L_t$, width $b$, and thickness $c$. The coordinate axes
are chosen as shown in Fig.~1. The semiconductor film with the
thickness $c/2$ occupies the length $L$ of the beam. The rest part
contains a metal film. It can also include some additional
elements for detection of the torque, for example in
Ref.~\cite{Mohanty}. Here we will consider an example when the
spin current is created by diffusion of the spin polarization from
the left contact in Fig.~\ref{system}. Therefore, there is no
electric current flow through NMS. The spin polarization diffuses
towards the metal film which, due to its relatively high
conduction, can play an important role as a reservoir for the spin
polarization relaxation.

We start from the strain-induced SOI~\cite{Soibystrain} described by
the Hamiltonian
\begin{eqnarray}\label{HSO1}
H_{\rm{SO1}}  &=& \alpha \left[\sigma _x \left(u_{zx}k_z-u_{xy}k_y\right)
+ \sigma _y\left(u_{xy}k_x-u_{yz}k_z\right)\right.\nonumber \\
&& + \left. \sigma _z \left(u_{yz} k_y-u_{zx} k_x\right)\right]  +
\beta \left[\sigma _x k_x \left(u_{yy}-u_{zz}\right)\right. \nonumber \\
&& + \left. \sigma _y k_y \left(u_{zz}  - u_{xx}\right) + \sigma _z
k_z\left(u_{xx}-u_{yy}\right)\right],
\end{eqnarray}
where $u_{ij}$ are elements of the strain tensor, $\sigma_i$ stand
for Pauli matrices and $k_i$ denote components of the electron wave
vector. In the narrow gap semiconductors the parameter $\beta$ is
usually much smaller than $\alpha$ \cite{Soibystrain}. Therefore,
the term proportional to $\beta$ will be omitted below. Besides the
strain-induced $H_{\rm{SO1}}$, the total SOI Hamiltonian also
includes the strain independent interaction $H_{\rm{SO2}}$. Because
of submicron cross-section dimensions of the doped semiconductor
film, $H_{\rm{SO2}}$ will be determined by the bulk Dresselhaus term
\cite{Dresselhaus}.
\begin{equation}\label{HSO2}
 H_{\rm{SO2}}=\delta \sum_{ijn}|\epsilon^{ijn}|k_i(k^2_j - k^2_n)
\end{equation}
This interaction, in the range of doping concentrations
10$^{17}$cm$^{-3}$ and higher, provides the main mechanism for
spin relaxation in bulk materials~\cite{Soibystrain}.

Since the S-M rod with total length $L_t \gg b$ and $c$, the major
contribution to the strain comes from flexural and torsion motions
of the rod~\cite{Landau}. Within the isotropic elastic model the
flexural motions are represented by the diagonal elements $u_{xx}$
and $u_{yy}$ \cite{Landau} which do not enter into the first square
brackets of \eq{HSO1}. On the other hand, due to the crystal
anisotropy effects, the $u_{xy}$ components are not zero for such
sort of motion and could contribute to \eq{HSO1}. We will consider,
however, the simplest example of torsion motions of the rod within
an isotropic elastic model. In this case the strain can be
represented as~\cite{Landau}
\begin{equation}\label{uyx}
u_{yx}=\tau(x)\frac{\partial \chi}{\partial z};\,\,\, u_{zx}= -\tau
(x)\frac{\partial \chi}{\partial y};\,\,\, u_{yz}=0 \,,
\end{equation}
where $\tau (x) = \partial \theta / \partial x$ stands for the
rate of torsion determined by the torsion angle $\theta$. The
function $\chi$ depends only on $z$ and $y$ and is uniquely
determined by the rod cross-section geometry.

The next step is to derive from the one-particle interaction
\eq{HSO1} a Hamiltonian which describes a coupling of the spin
current to the strain. The electron system carrying the spin current
can be described by a density matrix $\hat{\rho}$. In the framework
of the perturbation theory the leading correction to the electron
energy due to the SOI induced strain can be obtained by averaging
$H_{\rm{SO1}}$ with $\hat{\rho}$. In the semiclassical approximation
such a procedure can be represented as averaging over the classical
phase space with the Boltzmann distribution function
$\hat{F}_{\mathbf{k}}(\mathbf{r})$. This function is a 2$\times$2
matrix in the spinor space. One can also define the spin
distribution function $P^i_{\mathbf{k}}(\mathbf{r})=(1/2)
Tr[\hat{F}_{\mathbf{k}}(\mathbf{r})\sigma^i]$. It is normalized in
such a way that the local spin polarization
$P^i(\mathbf{r})=\sum_{\mathbf{k}}\mathbf{P}_{\mathbf{k}}(\mathbf{r})$
We notice that, due to electron confinement in $y$ and $z$
directions, the averages of $H_{\rm{SO1}}$ containing $k_y$ and
$k_z$  turn to zero. Assuming that electron distribution is uniform
within the cross-section of the semiconductor film one thus obtains,
from Eqs. (\ref{HSO1}) and (\ref{uyx}), the  SOI energy
\begin{eqnarray}\label{ESO}
E_{\rm SO}&=&2\alpha\int_0^L dx \frac{\partial \theta }{\partial
x} \nonumber
\\
&\times& \sum_{\mathbf{k}} k_x \int dydz
\left(P^y_{\mathbf{k}}(x)\frac{\partial \chi}{\partial z} +
P^z_{\mathbf{k}}(x)\frac{\partial \chi}{\partial y}\right)\, .
\end{eqnarray}
This expression can be further simplified taking into account that
$\chi$ turns to zero on a free surface \cite{Landau}. Hence, in the
example under consideration $\chi=0$ on the top and side surfaces of
the doped semiconductor film. Consequently, the second term in
\eq{ESO} vanishes after integration over $y$. Now \eq{ESO} can be
expressed in terms of the spin current $J^y(x)$ which is the flux in
$x$-direction of $y$-polarized spins.
\begin{equation}\label{current}
J^y(x)  = S \sum\limits_\mathbf{k} v_{x} P^y_{\mathbf{k}}(x) \, ,
\end{equation}
where $S=bc/2$ is the semiconductor film cross-section and $v_x$ is
the electron velocity in $x$-direction \cite{comment}. Finally,
\eq{ESO} can be transformed to
\begin{equation}\label{ESO2}
E_{\rm{SO}} = \gamma \int_0^L dx J^y(x)\frac{{\partial \theta
}}{{\partial x}}\, .
\end{equation}
Here the coupling constant $\gamma$ is given by
\begin{equation}\label{gamma}
\gamma  = \gamma_0 \int_{ - {\textstyle{b \over 2}}}^{{\textstyle{b
\over 2}}} \chi \left( {y,z=0}\right) dy \, ,
\end{equation}
where $\gamma_0 = 2m^*\alpha / \hbar S$.

From the last equation, it is seen that the spin-polarized flow
imposes a distributed torque on the rod. In order to study this
effect in detail we will neglect, for simplicity, the difference
between elastic constants of semiconductor and metal parts of NMS.
As such, the equation of motion for the torsion angle can be then
written as
\begin{equation}\label{eqmotion}
I\frac{{\partial ^2 \theta }}{{\partial t^2 }} - K\frac{{\partial ^2
\theta }}{{\partial x^2 }} - \gamma \frac{{\partial }}{{\partial
x}}\left[J^y\eta(L-x)\right] = 0\, ,
\end{equation}
where $\eta(x)$ denotes the Heaviside function, $K$ stands for the
torsion rigidity, and $I$ is the moment of inertia.  It is easy to
figure out that the torque imposed by the SOI on NMS can be
expressed as
\begin{equation}\label{torque}
\mathcal{T}=\frac{\gamma}{L} \int_0^L dx J^y(x) \equiv \gamma
\bar{J}^y\, ,
\end{equation}
and, for the S-M rod clamped on both ends, the torsion angle at
$x=L$
\begin{equation}\label{angle}
\theta_L=\frac{L(L_t-L)}{L_t}\frac{\mathcal{T}}{K} \, ,
\end{equation}
where $L_t$ is the total length of the rod. From \eq{eqmotion} one
can easily see that if the semiconductor film covers the entire
length of the beam ($L=L_t$) and the spin current is homogeneous
along it, the last term in \eq{eqmotion} turns to 0. Consequently,
for a doubly clamped beam the solution of \eq{eqmotion} is
$\theta(x)\equiv 0$. In this case, in order to obtain the finite
torsion angle, the NMS must include films with different spin-orbit
coupling parameters $\gamma$, as in Fig.~1 where $\gamma=0$ in the
metal film. On the other hand, if $J^y$ depends on $x$, as in the
example considered below, the metal film is not so necessary. In
this example it is shown, however, that such a film can be useful as
a reservoir for fast spin relaxation, enhancing thus the diffusive
spin current flow through the beam.

In order to evaluate the torque, let us adopt the following simple
model, which is also convenient for an experimental realization.
Namely, we assume that the spin current is due to spin diffusion
from the left contact. The spin polarization $P^y(0)$ can be created
there by various methods ranging from absorption of circularly
polarized light to injection from a ferromagnet~\cite{Rev.Mod.Phys}.
One more possibility is the electric spin orientation~\cite{Kato}.
For the steady state the diffusion equation reads
\begin{equation}\label{diffusion}
D_i \frac{{d^2 P^y }}{{dx^2 }} - \frac{P^y}{\tau _i}   = 0\, ,
\end{equation}
where $D_i$ and $\tau_i$ are diffusion coefficients and spin
relaxation times, with the subscript $i$ indicating the physical
quantities in semiconductor $(0<x<L)$ (i=S) or metal $(x>L)$ (i=M)
regions. At the semiconductor-metal interface the diffusion current
and magnetization $P^y/N_i(0)$ must be continuous, where $N_i(0)$ is
the semiconductor or metal density of states at the Fermi
energy~\cite{Johnson}. We will assume that the length of the metal
part of the rod is larger than the spin diffusion length
$l_M=\sqrt{D_M\tau_M}$. Therefore, the spin current passes through
the semiconductor film and further decays within the metal film.
Obviously, in the considered example there is no charge current
through the system. Solving the diffusion equation for $l_S \gg L$
and $(\sigma_M L)/(\sigma_S l_M) \gg 1$, where $\sigma_M$ and
$\sigma_S$ are the 3D conductivities of metal and semiconductor,
respectively, we obtain
\begin{equation}\label{Jbar}
\bar{J}^y=\frac{D_S P^y(0) S}{L}\,.
\end{equation}
Since the ratio $\sigma_M /\sigma_S$ is very big, \eq{Jbar} is valid
in a broad range of not very small $L$.

For a numerical evaluation of the spin-orbit torsion effect we take
$b=400$ nm and $c=200$ nm. The SOI coupling constant
$\alpha/\hbar=4\times 10^{5}$ m/sec in GaAs \cite{Dyakonov}. From
\eq{gamma} and Ref.~\cite{Landau}, it is easily to obtain the
spin-current--torsion coupling parameter $\gamma = \gamma_0 k_2
b^3$, where $k_2$ is a numerical factor depending on the ratio
$c/b$. At $b/c=2$ the factor $k_2=0.03$. For such numerical
parameters we find $\gamma=2.4\times 10^{-32}$ J sec. It is
interesting to compare the torsion effect from the strain induced
SOI with that produced by spin flips at the FM-NM
interface~\cite{Mohanty}. In the latter case $\mathcal{T}=\hbar
I_s$, where $I_s$ is of the order of the spin current injected at
the FM-NM contact when the electric current passes through it.
Comparing this expression with \eq{torque}, it is seen that at the
\emph{same} spin currents the SOI effect is much stronger, by the
factor $\gamma/\hbar\simeq 2.2\times 10^2$. On the other hand,
in~\cite{Mohanty} the FM-NM contact can be fabricated from all
metallic components, while our device must contain the narrow gap
semiconductor film. In the former case NMS is able to carry much
larger spin current, due to the weaker, by the factor $\sim
\sigma_S/\sigma_M$, Joule heating effect. However, the measurement
setup suggested by Mohanty \emph{et. al}~\cite{Mohanty} allows to
measure torsion effects produced by quite weak currents. For
example, at $e\bar{J}^y=10^{-8}$Amp  the torque
$\mathcal{T}=1.5\times 10^{-21}$ N m, which is within the
sensitivity claimed in~\cite{Mohanty}. Moreover, the measurement
sensitivity can be enhanced~\cite{Mohanty_pc}. Within our model we
can evaluate the spin polarization $P^y(0)$ which can produce a
measurable effect on NMS. From \eq{Jbar}, taking $L=2\, \mu{\rm m}$,
the typical low temperature diffusion constant 300 cm$^2$/sec and
$n=10^{17}$ cm$^{-3}$, one obtains $e\bar{J}^y=10^4 (P^y(0)/n)$ nA.
Hence, a measurable $10$ nA spin current in NMS can be created by
diffusion of spin polarization from an adjacent reservoir containing
only $0.1 \%$ of spin-polarized carriers. Various
methods~\cite{Rev.Mod.Phys,Wunderlich,Kato} are able to provide such
and even much larger spin polarization. Higher spin currents are,
however, restricted by the heating effects, which depend on the
practical design of NMS.

It should be noted that the torsion measurement method of
Ref.~\cite{Mohanty} applies to a time-dependent torque in
resonance with a NMS oscillation. For such a measurement the spin
current could be modulated in time by a narrow gate between the
left contact and the rod, or by varying the spin polarization in
the left reservoir, for example, if it is created by absorption of
circularly polarized light with modulated intensity.

The static torsion angle at $x=L$ can be found from \eq{angle}. On
the other hand, the maximum torsion effect is obtained for the
time-dependent spin current in resonance with the NMS fundamental
oscillation. In this case, the torsion angle $\theta_L$ in
\eq{angle} must be multiplied by $Q/2$, where $Q$ is the resonance
quality factor, which can be quite large in NMS.  To observe this
torsion angle it must be much larger than the mean amplitude of its
thermal fluctuations $\sqrt{\bar{\delta\theta^2_L}}$. For a doubly
clamped rod
\begin{equation}\label{fluctuation}
\bar{\delta\theta^2_L}=\frac{k_BT L_t}{\pi^2 K}\sum_{n\geq 1}
\frac{1}{n^2}\sin^2\left(\frac{\pi n L}{L_t}\right).
\end{equation}
For a rectangular cross-section with $b/c=2$, the torsion rigidity
$K=0.057\mu b^3c$~\cite{Landau}, where $\mu\simeq 3.3\times 10^{10}
\,{\rm N/m}^2$ in GaAs material. Taking $L_t=5\, \mu{\rm m}$ and all
other parameters the same as in the previous paragraph, $Q=10^4$ and
$T=100$ mK we obtain the ratio $\delta\theta_L/\theta_L \simeq
4\times 10^{-2}$ at $e\bar{J}^y=10$ nA.

We have considered a simple example of the spin-orbit torque effect
produced by spin flux in a diffusive 3D semiconductor film. It would
be interesting to study other systems, for example, a superlattice
of remotely doped high mobility quantum wells in the ballistic
regime ($L$ is less than the elastic mean free path). In such a
system energy dissipation within the semiconductor film is reduced
and, apparently, larger spin currents are allowable.

In summary, we propose a nanomechanical system where due to the
strain-induced spin-orbit interaction the electron spin degree of
freedom can couple to NMS mechanical motions. We have shown that
this coupling is strong enough to induce the measurable torsion in
NMS when the spin polarization flows through the suspended nanobeam.
Besides a potential for other possible applications, such NMS can be
employed as a sensitive detector of spin currents and spin
polarizations. The basic structure can be further modified to create
devices for eventual use in spintronics as well as spin information
processing.\\

This work was partly funded by the Taiwan National Science Council;
and RFBR grant No. 03-02-17452.



\begin{thebibliography}{99}
\bibitem{comment}
We neglected the small correction to the spin current associated
with Dresselhaus's term $\frac{\partial H_{\rm{SO2}}}{\partial
\mathbf{k}}$ in the velocity operator, as well as the corresponding
correction due to $H_{\rm{SO1}}$, within the linear perturbation
theory on the strain induced SOI.
\end{thebibliography}
\end{document}